\documentclass[preprint,showpacs,amsmath,amssymb,aps,pra]{revtex4-1}
\usepackage{graphicx}% Include figure files
\usepackage{natbib}
\usepackage{blindtext}
\usepackage{dcolumn}% Align table columns on decimal point
\usepackage{ulem}
\usepackage{bm}% bold math
\usepackage{amsmath}
\usepackage[utf8]{inputenc}
\usepackage{hyperref}
\usepackage{balance}
\usepackage{multirow}
\usepackage{longtable}
\usepackage{epstopdf}
\usepackage{comment}
\usepackage{xcolor}
\usepackage{dcolumn}% Align table columns on decimal point
\renewcommand{\k}{{\bbox k}}
\renewcommand{\k}{{\bm k}}

\newcommand{\be}{\begin{equation}}
\newcommand{\ee}{\end{equation}}
\newcommand{\br}{\begin{eqnarray*}}
\newcommand{\er}{\end{eqnarray*}}
\newcommand{\ba}{\begin{eqnarray}}
\newcommand{\ea}{\end{eqnarray}}
\newcommand{\bp}{\begin{minipage}}
\newcommand{\ep}{\end{minipage}}

\newcommand{\bt}{\begin{tabular}}
\newcommand{\et}{\end{tabular}}

\begin{document}
%\title{Zeptosecond dynamics in atoms: fact or fiction?}
\title{Zeptosecond to attosecond dynamics in atoms and possibility of generating a zeptosecond light source}
\author{T. Nandi$^{1*}$, Soumya Chatterjee$^2$, Adya P. Mishra$^3$, Y. Azuma $^4$, F. Koike$^5$ and A.S. Kheifets$^6$} %V. S. Yakovlev$^5$,}
\affiliation{$^{1}$ Department of Physics, Ramakrishna Mission Vivekananda Educational and Research Institute, PO Belur Math, Dist Howrah 711202, West Bengal, India.}
%\thanks {Email:\hspace{0.0cm} nanditapan@gmail.com} %Present address: 1003 Regal, Mapsko Royal Ville, Sector-82, Gurgaon-122004, India.}
\affiliation{$^2$Department of Physics, Brainware University, Barasat, Kolkata-700125, WB, India.}
\affiliation{$^3$Atomic \& Molecular Physics Division, Bhabha Atomic Research Centre, Trombay, Mumbai - 400 085, India}
\affiliation{$^4$Department of Physics, Indian Institute of Technology, New Delhi, Delhi-110016, India.}
\affiliation{$^5$ Sophia University, 7-1 Kioichō, Chiyoda City, Tokyo 102-8554, Japan}
\affiliation{$^{6\dagger}$Research School of Physics and Engineering, The Australian National University, Canberra,  ACT 0200, Australia }
%\thanks {Email:\hspace{0.0cm}A.Kheifets@anu.edu.au}
%%%%%%%%%%%%%

%\end{document}
%
\begin{abstract}
In nuclear collisions,  nuclear bremsstrahlung can cause nuclear Coulomb excitation via photon exchange in the projectile as well as  the target nuclei. Such a process originating in nuclear timescales (zeptoseconds) can also influence the atomic phenomenon, which can be observed if it is delayed at least by a few attoseconds as atomic timescales $\ge$ an attosecond. We have found that this may happen due to a mechanism called the Eisenbud-Wigner-Smith (EWS)
time delay process. We have estimated EWS time delays in atomic collisions utilizing the non-relativistic version of random phase approximation with exchange as well as Hartree-Fock methods. We present three representative collision systems through  which one can experimentally observe the phenomena in attosecond timescales even though they originate from nuclear bremsstrahlung radiation occurring in zeptoseconds. Thus the present work represents an investigation of parallels between two neighboring areas of physics: atomic and nuclear physics. Furthermore the present work suggests the possibilities for atomic physics research near the Coulomb barrier energies, where the nuclear bremsstrahlung can be used as a zeptosecond x-ray source.
\end{abstract}
\maketitle
%%
%\pacs{25.70.Bc, 25.70.-z, 25.70.Lm, 34.50.Fa}
%%
\maketitle
\section{Introduction}\label{intro}
During inelastic collisions of charged projectiles with atoms, atomic Coulomb and Pauli excitation \cite{basbas1974universal} cause inner-shell vacancy production through direct ionization to the continuum of the target atoms or by electron capture from the target atoms into an unoccupied state of the projectile ions. This process, discovered in the 1930s, is known as Coulomb ionization (CI) of inner shells by heavy charged particles \cite{brandt1974binding,brandt1979shell}. It is a typical atomic or electromagnetic process having larger range and smaller coupling constant than the corresponding quantities of the strong nuclear force. Thus the nuclear phenomena usually do not have any influence on this or any other atomic processes. However, in early fifties, an experiment detected K x-rays accompanying  $\alpha$ decay process from radioactive $^{210}$Po \cite{barber1952evidence} and this x-ray emission phenomena would not be described by considering only the Coulomb ionization process \cite{ciocchetti1965k}. They suggested a coincidence experiment to observe it more discerningly.
%%%%%%%%%%%%%%%%%%%%%%%%%%%%%
\par 
In late 1970s, Blair et al. \citep{blair1978nuclear} succeeded in observing the influence of nuclear process on atomic phenomenon for the first time through a clear rise in the Ni K x-ray production cross-section measured in coincidence with elastically scattered protons, while its energy was passed over the s$_{1/2}$ nuclear resonance at 3.151 MeV. Subsequently, a few more such experimental evidences on enhanced K-shell ionization were found due to the s$_{1/2}$ nuclear resonance at 461 keV of $^{12}$C \cite{duinker1980experimental} and at 5.060 MeV of $^{88}$Sr \cite{chemin1981measurement}. A theoretical study \cite{blair1982theory} suggested that a monopole excitation might be responsible in exhibiting the influence of a nuclear resonance on enhancing K-shell ionization provided the resonance width is less than or equal to the K- shell binding energy or equivalently, the time delay must not be less than the K-shell orbiting period. On the other hand, Greenberg et al. \cite{greenberg1977impact} observed large enhancement of the K-shell ionization cross-section at small impact parameters in a heavy-ion collision experiment at energies above the Coulomb barrier. This fact was interpreted as a contribution of the nuclear rotational-coupling mechanism  in addition to radial coupling proposed by Betz {\it et. al.} \cite{betz1976direct}.
\par 
About 40 years later in 2017 Sharma and Nandi \cite{Prashant-PRL-2017} have observed unusual resonance-like structures in the K x-ray spectra as the beam energy approaches the fusion barrier energy. The resonance structures were observed in the K x-ray energy of the elastically scattered projectile ion spectrum versus the beam energy plot \cite{Prashant-PRL-2017}. Such resonances have been attributed to the shakeoff ionization due to sudden projectile nuclear recoil due to nuclear force. Note that projectile x-ray energy corresponds to the mean charge state of the projectile ions inside the target foil and the higher x-ray energy implies the higher charge states \cite{sharma2016x}. Thus, variation of K x-ray energy is nothing but the variation of mean charge state of the projectile ions with the beam energy and  the same resonance must be reflected in the mean charge state versus beam-energy too. Though the nuclear recoil induced shakeoff mechanism explains well the measured variation of the mean charge $q_m$ including the resonance structure \cite{Prashant-PRL-2017}, it is our curiosity to check whether a bremsstrahlung radiation evolved due to the retardation of the projectile ions by the nuclear force, called nuclear bremsstrahlung  \cite{jakubassa1975bremsstrahlung}, can also elucidate the said resonance structure too. The basis of  the idea  is germinated because the phenomena such as nuclear resonances \cite{griffy1962nuclear}, nuclear Coulomb excitation \cite{alder1956study}, nuclear giant dipole resonances \cite{izumoto1981coulomb}, etc are originated from the nuclear bremsstrahlung radiations. These nuclear processes occur in timescales of the order of zeptoseconds ($zs$), whereas, the x-ray phenomena are atomic events, which take place in the timescales  of the order of attoseconds ($as$). In this paper, we make an attempt to address an intriguing question: how a $zs$ event initiates an $as$ phenomena.
%%%%%%%%%%%%%%%%%%%
\section{Significance of Lienard-Wiechert potentials in atomic collisions}\label{L-W pot}
%Since the nuclear bremsstrahlung is caused by retardation of the projectile due to nuclear force and the retardation is a result of nuclear recoil, the nuclear bremsstrahlung is the total effect whereas the nuclear recoil is the initial stage of the process. \\
%%%%%%%%%%%%%%%%%%%%%
The nuclear Coulomb excitation takes place during the heavy-ion collisions from intermediate \cite{bertulani2003intermediate} to relativistic energies \cite{wollersheim2011relativistic}. The origin of the nuclear Coulomb excitation is as follows: While the projectile approaches the target nucleus, it faces the nuclear interaction barrier potential (B$_{int}$) \cite{basbas1974universal} and thus it is retarded and its velocity is reduced considerably. For example, for the $^{63}$Cu projectiles having ion velocity $v=6.351({E_{p}/A_p})^{(0.5)}=9.6$~a.u., it is reduced to re$v'=6.351({E_{ret}/A_p})^{(0.5)}=$ 3.4 a.u., where $E_{ret}$ = retarded energy=$E_{res}-B_{int}$. Here, $E_{p}$ is the kinetic energy of the projectile ions. $E_{res}$ is  the resonance energy, which is measured experimentally. It appears at the projectile energy where the mean charge state takes a sudden jump \cite{Prashant-PRL-2017}). $B_{int}$ is the interaction barrier, which is estimated from an empirical formula given by \citet{nandi2022search}. The $A_p$ is the mass of the projectile ion.  Hence, the change in projectile velocity $\Delta v=(v-v')\times$ Bohr velocity due to influence of nuclear interaction is $1.3\times10^7$ m/s. This deceleration results in emission of electromagnetic radiation, which can be described by using the Lienard-Wiechert potentials \cite{griffiths2005introduction}. Such  electromagnetic radiations in an inelastic collision produce excitation in the interacting nuclei \cite{alder1956study,bertulani2003intermediate,wollersheim2011relativistic}. The origin of Coulomb excitation is associated with the power radiated due to deceleration of the projectiles by the target nuclei.
\begin{figure}[!h]
  \centering
   \includegraphics[width=8.5cm, height=4.0cm]{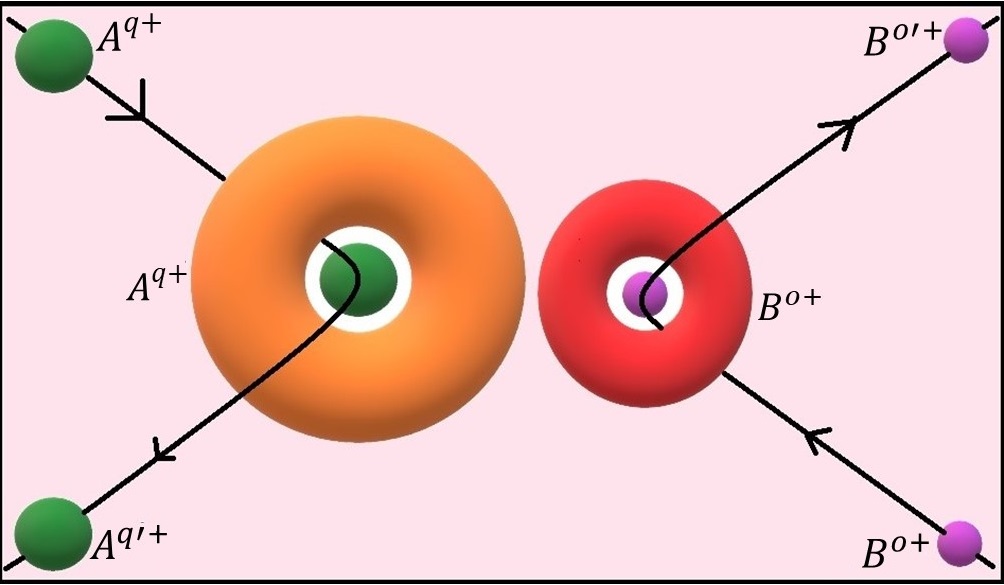}
    \caption{Schematic of the power radiated in the centre of mass frame due to deceleration by a point charge (projectile ion). The 
    charge state of the projectile ion is changed due to the deceleration and the same happens for the target ions too. Here, A$^{q+}$ denotes the projectile ions and B$^{o+}$ the target ions.}
    \label{fig:scattering}
\end{figure}
%%%%%%%%%%%%%%%%%%%%%%%%%%%%%%%%%%%%%%%%%%%%%%%%%%%
%\par
A schematic of the above mentioned retardation mechanism is shown in Fig.\ref{fig:scattering}. The power is radiated in a doughnut about the direction of   deceleration of the projectile or target nuclei, not in the forward or backward direction. It can be absorbed by the projectile ion or the target atom. The total power radiated ($P$) by the projectile corresponding to nuclear bremsstrahlung radiation can  be calculated from the Larmor formula \cite{griffiths2005introduction} in the center of mass frame as follows
\begin{equation}
P = \mu_0~\frac{q^2 f^2}{6\pi c}~\hspace{10pt}\text{(SI units)}\hspace{10pt}  
\label{Poynting}
\end{equation} 
\noindent Here $\mu_0$ is the permeability in vacuum, $q$ is the nuclear charge of the projectile, $c$ is the velocity of light, and $f$ is the deceleration of the projectile on encountering the target nucleus, which is obtained by the change in the velocity of the projectile occurring in collision time.  
%%%%%%%%%%%%%%%%%%%%%%%%%%%
\section{Estimating the collision time and total radiative energy}\label{coll time}
The calculation of the collision time $\tau_{coll}$ is important to estimate the total radiative energy. According to \citet{alder1956study}, the two limits of $\tau_{coll}$ i.e., the maximum collision time $\tau^{max}_{coll}$ and the minimum collision time $\tau^{min}_{coll}$ are given by
\begin{equation}
\tau^{max}_{coll} = \frac{a}{v} ~~~\text{and} ~~~ \tau^{min}_{coll}=\frac{a\sin{(\theta_{cm}/2)}}{v}.
\label{Coll_time}s
\end{equation} 
\noindent Here the symmetrized parameter $ a = Z_1Z_2e^2/(m_0 v^2)$, where $m_0$ is the reduced mass of the target-projectile system, $v$ is the projectile  velocity as mentioned earlier, and $\theta_{cm}$ is the grazing angles. Since $f$ varies with inverse squared of the collision time, $\tau^{max}_{coll}$ causes the minimum power, whereas $\tau^{min}_{coll}$ describes the maximum.  At fusion barrier and sub-barrier energies, $\theta_{cm}$ is close to 180$^\circ$, hence we get $\tau^{min}_{coll}=\tau^{max}_{coll}$ at the sub-barrier energies. This condition holds well for the resonance energy also, which appears in between the interaction barrier and the fusion barrier energy. Note that both the barriers are estimated from an empirical formula \cite{nandi2022search}. Around the barriers total energy radiated from the projectile system E$_{rad}$ can be estimated as follows
\begin{equation}
E_{rad}=P\times\tau^{max}_{coll}.
\end{equation}
The quantities $E_{res}$, $B_{int}$, $E_{ret}$, $\tau^{max}_{coll}$, and  $E_{rad}$ are listed in Table I for the case of three reactions ($^{56}$Fe, ${58}$Ni and ${63}$Cu ions on $^{12}$C target). 
%%%%%%%%%%%%%%%%%%%%%%%%%%%%%%%%%%%%%%%%%%
\section{Significance of nuclear bremsstrahlung process}\label{significance}
The bremsstrahlung radiation is emitted in the form of doughnut shape about the direction of deceleration of the projectile as shown in Fig.\ref{fig:scattering}.  A part of the $E_{rad}$ will be absorbed by the projectile electrons. Magnitude of this part depends on the distance between the atomic centre and the doughnut centre. When these two centres coincide with each other then the absorbed power is maximum and such a situation is attained at the resonance energies. To release this largest absorbed energy, the projectile ion will undergo Auger cascade that can produce higher ionic states in the projectile ions than that produced from the CI process. This is the reason for the observation of a sudden increase in the projectile charge states at the resonance energy.
\par
At the resonance energy the deceleration of the projectile ion is maximum, consequently the absorbed energy is also maximum. Therefore, the frequency of the radiation is maximum. Furthermore, nuclear bremsstrahlung radiation in a doughnut shape about the direction of deceleration of the projectile is ought to be coherent. 
%%%%%%%%%%%%%%%%%%%%%%%%%%%%
\begin{table}
\caption{Total power radiated by the projectile ion due to its retardation by interaction barrier.
Reactions are specified by projectile (Proj.), target (Targ.),  resonance lab energy E$_{res}$ in MeV, and interaction barrier B$_{int}$ in MeV. We list retarded energy of the projectile $E_{ret}$ in MeV occurring in collision time $\tau^{max}_{coll}$ ($zs$), and total energy radiated E$_{rad}$ (keV) in this duration.}
%%%%%%%%%%%%%%%%%%%%%%%%%%%%%%%%%%%%%
\begin{tabular}{|c|c|c|c|c|c|c|c|}
\hline
Proj. & Targ. & E$_{res}$& B$_{int}$&E$_{ret}$& $\tau^{max}_{coll}$& E$_{rad}$\\
  &  & (MeV)& (MeV)&(MeV)& (zs) & (keV) \\\hline
 %\begin{tabular}[c]{@{}c@{}}E$_{abs}$\\(keV)\end{tabular}\\ \hline
 $^{56}$Fe    & $^{12}$C & 120 & 98.8  & 21.2  & 2.83 &49.6\\\hline 
$^{58}$Ni    & $^{12}$C & 134 & 113.8 & 20.4  & 2.50 & 56.2\\\hline 
$^{63}$Cu    & $^{12}$C & 143 & 125.3 & 17.7  & 2.63 &96.4\\\hline 
\end{tabular}
\end{table}
%%%%%%%%%%%%%%%%%%%%%%%%%%%%%%%%%%%%%%%%%%%%%%%%%%%%%%%%
\section{Projectile ionization due to Coulomb as well as Nuclear bremsstrahlung-induced ionization}\label{CI vs NBI}
\begin{figure}[!h]
  \centering
\includegraphics[width=\linewidth]{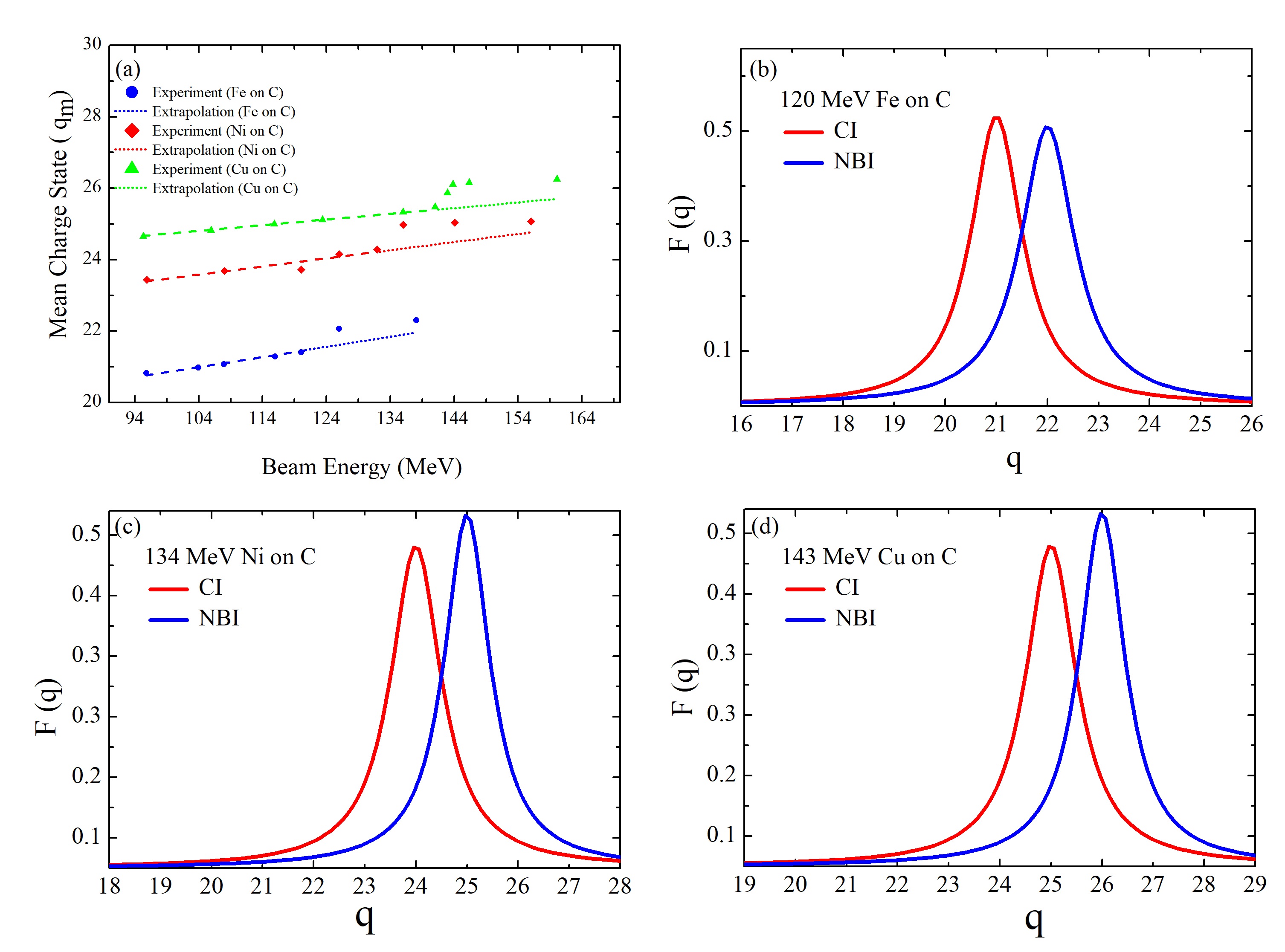}
\caption{Nuclear bremsstrahlung-induced ionization (NBI) versus Coulomb ionization (CI): (a) The symbols are the measured variation of q$_m$ with beam energy and the dashed lines represent straight line fit with the measured data. The extrapolation of the dashed lines for the energies $\ge$ E$_{res}$ is shown by the dotted lines indicates the measured variation of q$_m$ but in absence of the nuclear influence that occurs at energies $\ge$ E$_{res}$ \cite{Prashant-PRL-2017}, (b), (c) and (d) Charge state distribution due to CI (red) and NBI (blue) at the energies E$_{res}$ for $^{56}$Fe, $^{58}$Ni and $^{63}$Cu beam on $^{12}$C target, respectively.} % (c)  CSD due to CI and NBI for $^{58}$Ni beam on $^{12}$C target, and (d) CSD due to CI and NBI for $^{63}$Cu beam on $^{12}$C target. }
  \label{fig:CSD}
\end{figure}

For the projectiles of iron, nickel and copper, radiative and non-radiative channels are comparable. Thus, in the presence of multiple vacancies the x-ray emissions may occur along with the Auger electrons. Furthermore, multiple vacancies may originate Auger cascade process (similar to shakeoff process) instead of only the Auger transitions which can exhibit up to H- like lines. Since, collisions at E$_p \ge E_{res}$ contain multiple vacancies due to complete overlap of ionic centre and doughnut centre, as has been mentioned earlier, this may cause higher mean charge states. Hence, the mean charge states due to the NBI are expected to be higher than that caused by the CI, which has been exactly observed experimentally in Fig.\ref{fig:CSD}. The variation of the mean charge states ($q_m$) with beam energies has been displayed Fig. \ref{fig:CSD}(a). We can notice that $q_m$ varies quite smoothly till the occurrence of the resonance at a certain beam energy. This trend signifies that till the resonance point only the CI is responsible but from this point onward the role of nuclear bremsstrahlung supersedes the CI processes. If we fit the $q_m$ data up to the resonance energy with a straight line and extrapolate it to the higher energies, we get an idea how $q_m$ due to CI would have varied in the post resonance regime in the absence of the nuclear bremsstrahlung process. This extrapolation is shown by the dashed lines in Fig. \ref{fig:CSD}(a). Note that experimentally observed $q_m$ in the post resonance region is governed only by the Nuclear bremsstrahlung-induced ionization (NBI). This is justified by the fact that the effect of NBI  takes place in $zs$, which creates higher charged state than that produced by the CI. The role of CI that is effective in $as$ timescales cannot ionize the projectiles further because the ionization limit by the CI process has already been overtaken by the NBI process.
%%%%%%%%%%%%%%%%%%%%%%%%%%%%%%%%%%%%%
%[!h]
%%%%%%%%%%%%%%%
The $q_m$-values inside the target due to CI as well as NBI are obtained from X-ray spectroscopy experiments \cite{sharma2016x}. The variation of $q_m$ with beam energy is shown in Fig.2(a). We also know that the charge state inside the foil follows a Lorentzian distribution \cite{sharma2016x}. Thus, we can numerically obtain the charge state distribution (CSD) from the plot of charge state fraction $F(q)$ versus the charge state $q$. The  $F(q)$ is given by
\begin{equation}
F(q)=\frac{1}{\pi}\frac{\frac{\Gamma}{2}}{(q-q_m)^2+(\frac{\Gamma}{2})^2}  \label{CSF}
\end{equation}
\noindent 
Here distribution width $\Gamma$ is given by \citet{novikov2014methods}
%\begin{widetext}
\begin{equation}
\centering
\Gamma(x)= C[1-exp(-(x)^\alpha)][1-exp(-(1-x)^\beta)]\label{Gamma}
\end{equation}
\noindent where $x=q_m/z$, $\alpha=0.23$, $\beta=0.32$, and $C=2.669-0.0098\times Z_2+0.058\times Z_1+0.00048\times Z_1\times Z_2$. We note here that Eq. (4) works well for both the CI and NBI driven ionization processes. Here $q_m$ is the concerned parameter for the charge state distribution that makes difference between the CI and NBI in Fig. 2(b)-(d). For both the processes the $q_m$ is different at every beam energy, which differentiates the CI and NBI processes. The CI process works up to the resonance energy point and the NBI process acts from this point onward.
\noindent The distribution width $\Gamma$ and $q_m$ have the following  relationship  
\begin{equation}
\Gamma^{2}=[1-(\frac{q_m}{Z})^\frac{5}{3}]\frac{q_m}{4}. \label{Gamma1}
\end{equation}
The charge state fraction $F(q)$ always obeys the following condition
\begin{equation}
\sum_q F(q)=1. \label{fq}
\end{equation}

%%%%%%%%
We can see from Fig. 2(a) that $q_m$ is different for CI and NBI in the post resonance energy. The CSDs pertinent to each $q_m$ due to CI as well as NBI are shown in Fig. 2(b-d) for the experiments with iron, nickel and copper projectiles. One can see clearly that one unit higher charge states are produced due to the NBI from resonance energy onward than that by the CI. The NBI is the manifestation of the interaction of the projectile ions and the photons emanating from the nuclear bremsstrahlung process, whereas the CI results from the Coulomb interaction of the projectile ions and the target atoms.  %%%%%%%%%%%%%%%
\section{Timescales involved in the CI and NBI processes}\label{timescale}
The NBI originates from the nuclear process occurring in zeptoseconds and the CI from the atomic process occurring in attoseconds, hence, the NBI is much faster than the CI. If NBI causes the higher charge state in advance, the CI cannot have any role to play in this charge changing process. In terms of energy, the CI plays its role up to the resonance energy only, whereas the action of NBI starts at the resonance energy and continues to higher energies. Hence, in terms of the timescale as well as beam energy, the CI and NBI processes never overlap.
\par
To explain the difference between the nuclear and atomic timescales for the systems under consideration, we estimate the characteristic time ($t_0$) for the atomic states from the ratio of the expectation values of the corresponding electronic radius ($\langle{r}\rangle$) and velocity ($\langle{v}\rangle$), which turns out to be in attoseconds. According to the measured x-ray spectra \cite{Prashant-PRL-2017}, for the projectile energies in the experiments, the mean charge state at the exit channel corresponds to the Li-like ions. The $1s2s2p$~$^{2,4}P^o_{1/2,3/2,5/2}$ levels of  Li-like Fe, Ni, and Cu are mostly populated at the resonance energies. To evaluate $t_0$ for these levels, we computed $\langle{r}\rangle$ and $\langle{v}\rangle$ by the multi-configuration Dirac-Fock formalism using GRASP2K computer package \cite{jonsson2007grasp2k}. The values of $t_0$ for the three systems mentioned are found to be 0.383, 0.328, and 0.306 $as$, respectively, which are at least two orders of magnitude larger than the nuclear collision times ($zs$ timescale) when the NBI takes place. It thus raises a fundamental question: how does the NBI transcend into the atomic regime (in $as$ timescale)?%This fact is well explained below in terms of a theoretical calculation. 
\begin{figure}[!h]
  \centering
    \includegraphics[width=\linewidth]{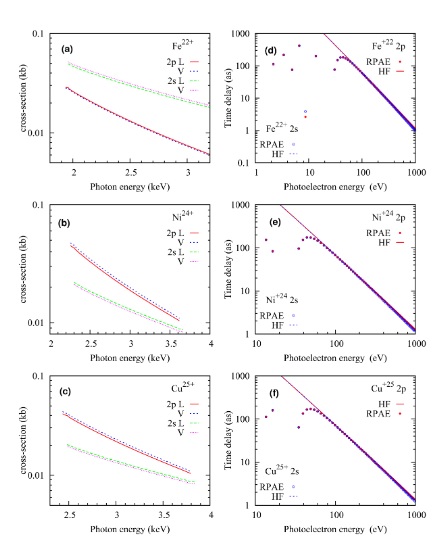}
  \caption{Photoionization cross-section as a function of photon energy of the emitted photons during the heavy-ion collisions shown in the left panel ((a), (b), and (c)).  The EWS time delay due to electromagnetic radiation induced photoionization versus photoelectron energy is shown in the right panel ((d), (e), and (f)). }
\label{timedelay}
\end{figure} 
\section{Estimating Eisenbud-Wigner-Smith time delay}\label{EWS delay}
The above mentioned radiative power may cause both the excitation and ionization in the ions in the short nuclear collision timescales ($zs$). The ionization will give rise to free electrons leaving the ion in the ground state, whereas the excitation can lead to autoionizing levels in the multielectron atomic systems. The autoionization process may decay through an excited state that relaxes into the ground state by emitting radiation with a higher x-ray energy, as shown in Fig. \ref{timedelay}. The autoionization process does not occur instantly as the electron has to move from the interaction regime to an interaction free regime, which introduces a time delay because of the difference between the density of states of the two regions \cite{ahmed2004number}. The energy-integral of time delay is an adiabatic invariant in quantum scattering theory and it provides a quantization condition for resonances \cite{jain2005}. In contrast, during CI both the excitation and ionization take place in the interaction zone. Thus, both of these two processes do not move from the field region to the field-free region and do not encounter any time delay.
\par
The time delay between different ionization channels, known as the Eisenbud-Wigner-Smith (EWS) time delay \cite{wigner55,smith60}, introduces finite delays. A delay of $21\pm 0.5 ~ as$ has been measured in a recent experiment of 100 eV light pulse induced on neon \cite{schultze2010delay}. This delay is interpreted as the difference in time delay of the emission of electrons liberated from the 2p orbitals of neon atoms with respect to those released from the 2s orbital by the same 100 eV light pulse. Hence, EWS time delay between the emission of electrons from different subshells in  various ionic states can take place too. Above experiment is different from the experiment considered in this work, but both are photon induced processes, one is driven by laser pulse and the other is caused by bremsstrahlung radiation. Thus excitation and ionization mechanisms are analogous to each other. In the present case also, the photoelectrons released from the 2s and 2p shells will differ similarly. To estimate this EWS time delay with the photoelectrons as well as the photoionization cross-sections, we have employed the non-relativistic versions of the random-phase approximation with exchange (RPAE) and Hartree-Fock (HF) methods \cite{ASK2015}. The partial photoionization cross-section for the transition from an occupied state $n_il_i$ to the photoelectron continuum state $kl$ is calculated as
\begin{equation}
\label{CS-HF} \sigma_{n_il_i\to kl}(\omega) = \frac43 \pi^2\alpha a_0^2\omega
\left| \langle k l\,\|\,\hat D\,\|n_il_i \rangle \right|^2 \,  
\end{equation} 
\noindent where $\omega$, $\alpha$ and $a_0$ are the photon energy,  the fine structure constant
 and the Bohr radius, respectively. All these quantities are used in atomic units $e=m=\hbar=1$.
In the independent electron Hartree-Fock (HF) approximation, the
reduced dipole matrix element is evaluated as a radial integral given below
\be
\label{reduced} \langle kl \|\,\hat D\,\|n_il_i\rangle=[l][l_i] \left(
\begin{array}{rrr} l&1&l_i\\ 0&0&0\\
\end{array}\right)\\\int r^2dr\, R_{kl}(r)\,r\,R_{n_il_i}(r)\,
\ee
where the notation $[l] = \sqrt{2l+1}$ is used. The basis of occupied atomic states $\|n_il_i\rangle$ is defined by the self-consistent HF method and calculated using the computer code
\cite{CCR76}. The continuum electron orbitals $\langle kl \|$ are defined
within the frozen-core HF approximation and evaluated using the
computer code described in \cite{CCR79}. 
\par
The present application of the RPAE rests on the electromagnetic dipole interaction between the projectile and the target. In this respect it is similar to the previous application in laser driven photoionization \cite{kheifets2013time}. The pole in the complex energy plane is explicitly accounted for by the RPAE, whereas the polarization of the target by the external field is not taken into account. This fact is very well justified in low wavelength laser-atom interaction. It is so in the fast and highly charged projectile interaction with the target atom also. In the RPAE, the reduced dipole matrix element is determined by summing an infinite sequence of Coulomb interactions between the photoelectron and the hole in the ionized shell. A virtual
excitation in the shell $j$ to the ionized electron state $\k'$ may
affect the final ionization channel from the shell $i$. This way the RPAE
accounts for the effect of inter-shell $i\leftrightarrow j$
correlation, also known as inter-channel coupling.  It is important to
note that, within the RPAE framework, the reduced dipole matrix element is 
complex and, thereby, adds to the phase of the dipole amplitude.
\par 
The photoelectron group delay, which is the energy derivative of the
phase of the complex photoionization amplitude, is evaluated as
\begin{equation}
\label{delay}
\tau = {d\over dE} \arg f(E)\equiv
{\rm Im} \Big[f'(E)/f(E) \Big].
\end{equation}
\noindent Here $f(E)$ is used as a shortcut for the amplitude 
$\langle\psi^{(-)}_\k|\hat z|\phi_i\rangle$ evaluated for $E=k^2/2$ and $\hat\k$ is $\parallel$ to $z$ axis, where $\psi^{(-)}_\k$ is an incoming scattering state with the given photoelectron momentum $\k$ \cite{A90}.\\
\par 
For the photoionization and time delay calculations, we have considered the charge species Fe$^{22+}$, Ni$^{24+}$, and Cu$^{25+}$ for the systems $^{56}$Fe + $^{12}$C, $^{58}$Ni + $^{12}$C, and $^{63}$Cu + $^{12}$C, respectively, as the Li-like ions are observed \cite{sharma2016x} after the autoionizaion. It is found that the EWS time delay, $\tau$, is proportional to  $\varepsilon ^{- 3/2} ln (1/\varepsilon$), where $\varepsilon$ is the photon energy. The elastic scattering phase, $\sigma$ ($\approx$ $\eta$ $ln {|\eta|}$, $\eta = -Z/\sqrt{2\varepsilon}$), is divergent near the threshold because of the Coulomb singularity \cite{ASK2015}. To remove this singularity, we have cut off the low energy photoelectrons for time delay calculations in the present computations. The results of the computations are displayed in  Fig.\ref{timedelay} and the  photoionization cross-sections are very close for the length ({\textit{L}}) and velocity ({\textit{V}}) gauges. %The photoionization cross-sections for the 2p electron in Ni$^{24+}$ and Cu$^{25+}$ are higher than that of 2s electron for all photon energies. Whereas, for Fe$^{22+}$, the  photoionization cross-sections for the 2s electron is higher than the 2p electron. 
%Further, the Wigner-Smith time delay  can transfer the  \rrt{orbiting  induced ionization } triggered  in the nuclear timescale ($zs$) to a few hundreds of $as$ for all the three systems considered here. Thereby, the photoionization phenomenon can occur in the atomic timescale and one measures the x-ray emissions. \\
%%%%%%%%%%
\section{Realization of the zeptosecond events in attosecond timescales}\label{zepto toatto}
Dissociation of the atomic excited states can occur only in timescales $\ge$ $as$,  because no atomic states exist with lifetime shorter than $as$. Hence, once an atomic observable such as the x-ray is observed, it proves that the phenomena occurs in attosecond timescales. Evolution of higher mean charge states than that occurred owing to CI is due to nuclear origin such as NBI and thus it originates in the zeptosecond timescale. This sort of occurrence is due to the emission of an electron from an excited state, which is delayed by $as$ according to the EWS time delay process or due to electron-electron correlation effects \cite{de2002time}. For example, photoemission of an electron is delayed up to six attoseconds in a helium atom \cite{ossiander2017attosecond}. Laser pulse induced photoionization of different states in neon are delayed differently. For the 2p state, it is delayed by 21$\pm5$ $as$ more than that for the 2s state in neon because of multi electron-correlation dynamics, as the s-electrons are ionized faster than the p-electrons \cite{schultze2010delay}. Once the absorbed photo-energy in projectile causes 1s and 2s shells ionized before the p-electrons, the projectile attains multiple innershell vacancies while any 2p-electrons are present in the ions and thus it results in Auger cascades. This picture has been theoretically reproduced  using the above mentioned approach and the same is applied for three test cases in the present work. 
%
%\begin{figure}
% \vspace{3mm}
%  \centering
 %     \includegraphics[width=9cm, height=10cm]{schematic.eps} 
 %\caption{Schematic of the overall x-ray emission mechanism at the resonance energies \cite{Prashant-PRL-2017} and beyond, where the atomic process is influenced by the nuclear bremsstrahlung process due to heavy-ion collisions at $\ge$ E$_{res}$ \cite{Prashant-PRL-2017}. See the text for details.}
 %   \label{fig:Schematic}
%\end{figure}
%
\par 
Let us consider the NBI that may result in Be-like ions and at least two electrons are in excited states. Naturally, the autoionization process of such Be-like ions will lead to the Li-like ions through a certain delay as discussed above.  This delay can be calculated as a function of the photo-electron energy using the Eqn. \ref{delay} as shown in Fig.\ref{timedelay}. One can see that the delays are in $as$ range for every photoelectron energy and the maximum delay is about 200 $as$ for the photoelectrons $\approx$ 60 eV. Furthermore, the figure displays an interesting feature that the emission of electrons liberated from the 2p orbitals is more delayed than that from the 2s-orbitals. This difference is about 20 $as$, which is in well accord with the measurements in Ne-atoms \cite{schultze2010delay}.
%%%%%%%%between the emission of electrons liberated from the 2p orbitals of neon atoms
\par 
%Effect of nuclear bremsstrahlung in the projectile ion at E$_p \ge$ E$_{res}$ is schematically described in %Fig.\ref{fig:Schematic}.
The projectile ion can be retarded as soon as it encounters an interaction barrier ($t=0$). Retardation is maximum at the saddle point/resonance energy, which gives rise to nuclear bremsstrahlung radiation from the projectile ion that can be absorbed in a doughnut shape containing the projectile electrons. This incident occurs during the short collision time in $zs$ duration as given in Table I. Total radiated energy absorbed, as estimated by the Larmor formula, is quite large and that can  be released through the simultaneous emission of many electrons. This is because a bunch of electrons get sufficient energy to be ionized and electron-electron correlation does not allow an individual electron to escape that leaves the projectile ion in much higher charged states (H- to O-like ions according to Fig. \ref{fig:CSD}). This photon induced ionization can take place at a duration larger (up to 200 $as$) than the atomic characteristic times (of the order of $as$) so that Blair and Anholt \cite{blair1982theory} criteria mentioned in the introduction section can be fulfilled. As a result, one can measure the $zs$ phenomenon using x-ray spectroscopy experiments that occurs in $as$ \cite{Prashant-PRL-2017}. \\ 
%%%%%%%%%%
\section{Nuclear recoil versus nuclear bremsstrahlung}\label{rec vs brems}
The increase in the mean charge state at the resonance energy has been explained using the nuclear recoil phenomenon \cite{Prashant-PRL-2017}. The resonance energy occurs at the sub-barrier energy. In this energy region the processes such as nuclear resonances \cite{griffy1962nuclear}, nuclear Coulomb excitation \cite{alder1956study}, nuclear giant dipole resonances \cite{izumoto1981coulomb} etc also take place. Nevertheless, nuclear recoil cannot originate such phenomena. This paper explains the increase in the mean charge state at the resonance energy by the nuclear bremsstrahlung process. It is also the cause of the nuclear resonances, nuclear Coulomb excitation, nuclear giant dipole resonances etc. The nuclear recoil phenomenon induces a shakeoff process that gives rise to the higher charge states of the projectile ions and the nuclear bremsstrahlung process initiates the Auger cascade that leads to the higher charge states of the projectile ions. The shake-off process and Auger cascade are very similar, both the processes cause the higher charge states. Hence, the present work infers that both the nuclear recoil and nuclear bremsstrahlung processes may be responsible for the observed sudden increase in the mean charge state at the resonance energy. However, evaluating the corresponding contribution of either one is far from our reach at this point of time. 
%%%%%%%%%%%%%%%%%%%%%%%%%
\section{Implications of the nuclear bremsstrahlung for atomic physics research}\label{implication}
The Nobel Prize in Physics 2023 \cite{dombi2023nobel} was awarded to Pierre Agostini, Ferenc Krausz and Anne L’Huillier for experimental methods that generate attosecond pulses of light for the study of electron dynamics in matter. However, the zeptosecond light source is far from reality till date. At this juncture, the present work provides an idea to make use of the nuclear bremsstrahlung as a zeptosecond light source by heavy ion impact with beam energies $E_p\ge E_{res}$ [see section \ref{significance}] where nuclear interaction begins to act. At such conditions the center of the projectile ions overlap to the doughnut center, hence efficiency of the nuclear bremsstrahlung interaction with the projectile ions is maximum. It means the absorption of radiated energy by the projectile ion is maximum from $E_p = E_{res}$ onward. The resonance occurs with the projectile ionization $E_p = E_{res}$ and the projectile ionization remains similar at any projectile energy $E_p > E_{res}$ too (see Fig.2(a)). The radiated energy amounts to tens of keV (see Table I) and it absorbs resonantly to the projectile ion. Hence, at the beam energy $E_p \ge E_{res}$, nuclear bremsstrahlung would produce the x-ray radiation of a very narrow frequency width. Using this zeptosecond narrow-bandwidth nuclear-bremsstrahlung x-ray source, we can study very well the atomic events occurring in $as$ including the time dependent properties,
especially light induced atomic ionization studies similar to the experiments of \citet{young2010femtosecond} using the femtosecond x-ray laser.
\par
The above mentioned fixed-frequency radiation with the nuclear bremsstrahlung interaction is possible  for a single collision in the target with a monochromatic ion beam. Since the beam energy has a certain spread and also multiple collisions take place in the target, the nuclear bremsstrahlung spectrum is broadened. As the targets used in nuclear physics experiments are normally thick, a large number of collisions takes place, hence the observed spectrum of the nuclear bremsstrahlung radiation is usually continuous. 
%%%%%%%%%%%%%%%%%%%%%%%%%%%
\section{Conclusion}
Coulomb ionization occurs at any projectile energy, but nuclear bremsstrahlung radiation induced ionization takes place in the vicinity of nuclear force i.e., when projectile energy is greater than or equal to the resonance energy (E$_p \ge$ E$_{res}$). Thus, the effect of  nuclear bremsstrahlung radiation induced ionization is  distinct as the Coulomb ionization occurs at every energy. Furthermore, the former originates when the center of projectile ion coincides with the doughnut center, which means that the ionizing source of radiation is sitting inside the atomic system. Whereas the latter is due to atomic collisions, in which the ionization is caused peripherally by the electronic excitation and electron stripping. Consequently, nuclear bremsstrahlung radiation can create multiple vacancies in the inner shells by ionizing the 1s and 2s shells prior to the 2p shells, which in turn causes the Auger cascades.  As a result NBI takes place in the short nuclear timescales i.e., of the order of zeptoseconds. In contrast, CI is a result of slow atomic collision processes occurring in $\ge$ an attosecond. The EWS time delay associated with the autoionization is found to be about 200 $as$ for three representative collision systems (Iron, nickel, and copper projectiles on carbon target). This delay is much larger than the lowest atomic timescale ($as$). Hence, we revealed that the events in the nuclear timescales can cause the phenomenon in the atomic timescales due to EWS time delay. 
\par
Till date the fastest light source achieved is attosecond pulsed laser \cite{drescher2001x}. Shorter than this is perhaps impossible as the limit of the atomic timescale is attoseconds. Though laser faster than attosecond is impossible, but making a light source of zeptosecond pulse width is possible by venturing nuclear time domain with a fast ion-beam. Such an idea is realized in the present article with appropriate experimental supports. We believe this article will inspire various important measurements using the zeptosecond light source for both atomic and nuclear physics.
%%%%%%%%%%%%%%%%%
%We show that nuclear bremsstrahlung can describe well the sudden jump of mean charge state at E$_p =$ E$_{res}$,  which is also cause of nuclear resonances, nuclear Coulomb excitation, nuclear giant dipole resonances etc. Hence, the nuclear bremsstrahlung is more appropriate phenomenon than the nuclear recoil as described by \cite{Prashant-PRL-2017} as the nuclear bremsstrahlung is the total effect whereas the nuclear recoil is the initial stage of the process. 
%%%%%%%%%%%%%%%%%%%
%%%%%%%%%%%%%%%%%%%%%%%%%%%%%%%%%%%%%%%%%%%
\section{Data availability}
All data generated or analyzed during this study are included in this published article.
%%%%%%%%%%%%%%%%%%%%%%%%%%%%%%%%%%%%%%%%%%%
\section{Acknowledgements}
We acknowledge useful discussions in various stages of this work with Prashant Sharma and help obtained from Gajendra Singh during preparation of the manuscript.
%Without which this work would not have come to this form.
\bibliographystyle{elsarticle-num-names}
\bibliography{WS.bib}
\end{document}